\documentclass[%
 aip,
 amsmath,amssymb,
 reprint,%
]{revtex4-1}

\usepackage{graphicx}
\usepackage{dcolumn}
\usepackage{bm}

\usepackage[utf8]{inputenc}
\usepackage[T1]{fontenc}
\usepackage{mathptmx}

\usepackage{xcolor}

\begin{document}

\preprint{APP20-LE-00202}

\title[Objective-free excitation of quantum emitters with a laser-written micro parabolic mirror]{Objective-free excitation of quantum emitters \\ with a laser-written micro parabolic mirror}

\author{Sergii Morozov}
 \altaffiliation[Present address:]{ Centre for Nano Optics, University of Southern Denmark, Campusvej 55, Odense M, DK-5230, Denmark.}
\author{Stefano Vezzoli}%
 \affiliation{%
The Blackett Laboratory, Department of Physics, Imperial College London, \\ London SW7~2BW, United Kingdom
}%

\author{Ali Hossain Khan}

\author{Iwan Moreels}
 \affiliation{%
Department of Chemistry, Ghent University, Krijgslaan 281-S3, Gent 9000, Belgium
}%

\author{Riccardo Sapienza}
\email{r.sapienza@imperial.ac.uk}
 \affiliation{%
The Blackett Laboratory, Department of Physics, Imperial College London, \\ London SW7~2BW, United Kingdom
}%

\date{\today}

\begin{abstract}
The efficient excitation of  quantum sources such as quantum dots or single molecules requires high NA optics which is often a challenge in cryogenics, or in ultrafast optics. Here we propose a 3.2~$\mu$m wide parabolic mirror, with a 0.8~$\mu$m focal length, fabricated by direct laser writing on CdSe/CdS colloidal quantum dots, capable of focusing the excitation light to a sub-wavelength spot and to extract the generated emission by collimating it into a narrow beam.
This mirror is fabricated via in-situ volumetric optical lithography, which can be aligned to individual emitters, and it can be easily adapted to other geometries beyond the paraboloid. This compact solid-state transducer from far-field to the emitter has important applications in objective-free quantum technologies. 
\end{abstract}

\maketitle

\section{\label{sec:level1}Introduction}

Highly confined optical fields are required for excitation of individual emitters and for efficient generation of single photons. Confining light to sub-wavelength volumes to decrease the background signal or to increase the fluency, is a common strategy in many optical domains, such as confocal microscopy, super-resolution microscopy, optical lithography, optical tweezers, ion and atom trapping, high density data storage, material processing.
Lens-based objectives are  traditionally used to focus light to a diffraction limited spot, a size which  depends on the numerical aperture (NA) \cite{Abbe1883}. High NA objectives are bulky and have to be operated at short working distances which complicates their practical applications, especially at cryogenic temperatures, while their chromatic aberrations   {and temporal dispersion hamper}  their use with ultra-short pulses.

Objectives can be replaced by reflective architectures which provide inherent achromaticity and  nonparaxial focusing due to the high NA. The simplest reflective objective is a parabolic mirror, which concentrates an incident beam around the geometrical focal point. The resulting focal spot has a size which is on a par with or better than that of high quality lens objectives \cite{Lieb2001,Stadler2008}.  Parabolic mirrors have found multiple applications in confocal microscopy \cite{Drechsler2001}, cryostat based single molecule spectroscopy \cite{Durand1999}, scanning optical near-field microscopy \cite{Sackrow2008}, Raman microscopy \cite{Zhang2009}, solar cells  \cite{Kosten2013}, light-emitting diodes \cite{Tanriseven2008}, nonlinear optics \cite{Penjweini2019}. 
Recently, deep parabolic mirrors  {covering $4\pi$ solid angle} gained attention in quantum optics because of their efficient focusing, ability of trapping individual quantum emitters, and of extracting a collimated beam of single photons \cite{Sondermann2015,Salakhutdinov2016}. 
However, future practical applications require miniaturisation and integration of parabolic mirrors.

When dealing with integrated photonics, different strategies to focus light into quantum emitters and collect their radiation have been explored. 
Nanofocusing has been achieved with structured metamaterials, as for example using a metalense based on plasmonic Fresnel plates \cite{Ma2010}, hypergratings \cite{Thongrattanasiri2009}, and plasmonic metamirrors \cite{Ding2019}. Plasmonic antennas have been shown to confine optical fields to zeptoliter volumes delivering nano resolution for plasmonic direct writing lithography \cite{Wang2016},  albeit often hampered by ohmic losses, difficulty in precise nanometric alignment with an emitter, as well as complicated and expensive fabrication techniques.

Deterministic integration of  micro optical components and individual quantum emitters can be achieved either by pre- or post-fabrication alignment, combined with lithographic fabrication. 
By lithographic techniques, various compact optical systems have been fabricated to control and manipulate light at the nanoscale, as for example waveguides \cite{Shi2016,Colautti2020}, polarisation rotators \cite{Schumann2014}, microdisc resonators \cite{Schell2013}, objectives \cite{Fischbach2017,Gissibl2016}, dielectric pillar antennas \cite{Au2019}, pillar microcavities \cite{Dousse2008}.  {High-index solid immersion lenses have been used to improve the coupling to quantum light sources, decreasing the laser excitation spot by a factor $1/n$ and magnifying the photoluminescence image of an emitter by a factor $n$ \cite{Sapienza2015,Sartison2017,Schmidt2019}. Microscale parabolic antennas have been shown to be an ideal design for directing light from quantum emitters \cite{Schell2014,Morozov2018}, however the focusing abilities of such compact structures have not been explored. }

In this letter, we report a compact parabolic mirror for sub-wavelength excitation of quantum emitters placed in its focal spot. The mirror also directs the generated photons into a low divergent beam along the parabola symmetry axis. The parabolic mirror is fabricated by in-situ optical volumetric lithography, which produces paraboloid structures in a single laser exposure step and results in high optical quality surfaces. We experimentally demonstrate that this mirror can focus light to a spot which is comparable to one of a high NA oil immersion objective. With a focal length of 0.8~$\mu$m, and a predicted focal spot of 120~nm ($\sim \lambda_{ex}/4$), the micro-mirror acts as an ideal optical transducer from the emitter to free-space.

\section{Results and Discussion}

\subsection{Focusing with a micro parabolic mirror}

A parabolic mirror illuminated with a collimated laser beam concentrates the excitation energy in its focal point. 
If the NA is large enough, the mirror has a sub-wavelength focal spot formed with minimal aberrations \cite{Lieb2001}, and broadband response in the visible and near infrared range of electromagnetic spectrum. 
 {While refractive optics is intrinsically limited by the frequency-dependent dielectric constant of its constituent, this is not the case for the parabolic mirror which is intrinsically achromatic (i.e. different wavelengths focus always in the same plane). Instead, the different refractive index experienced by wavelength will affect the focal spot size (see SI Fig.S2).}

\begin{figure}
\includegraphics[width=1\linewidth]{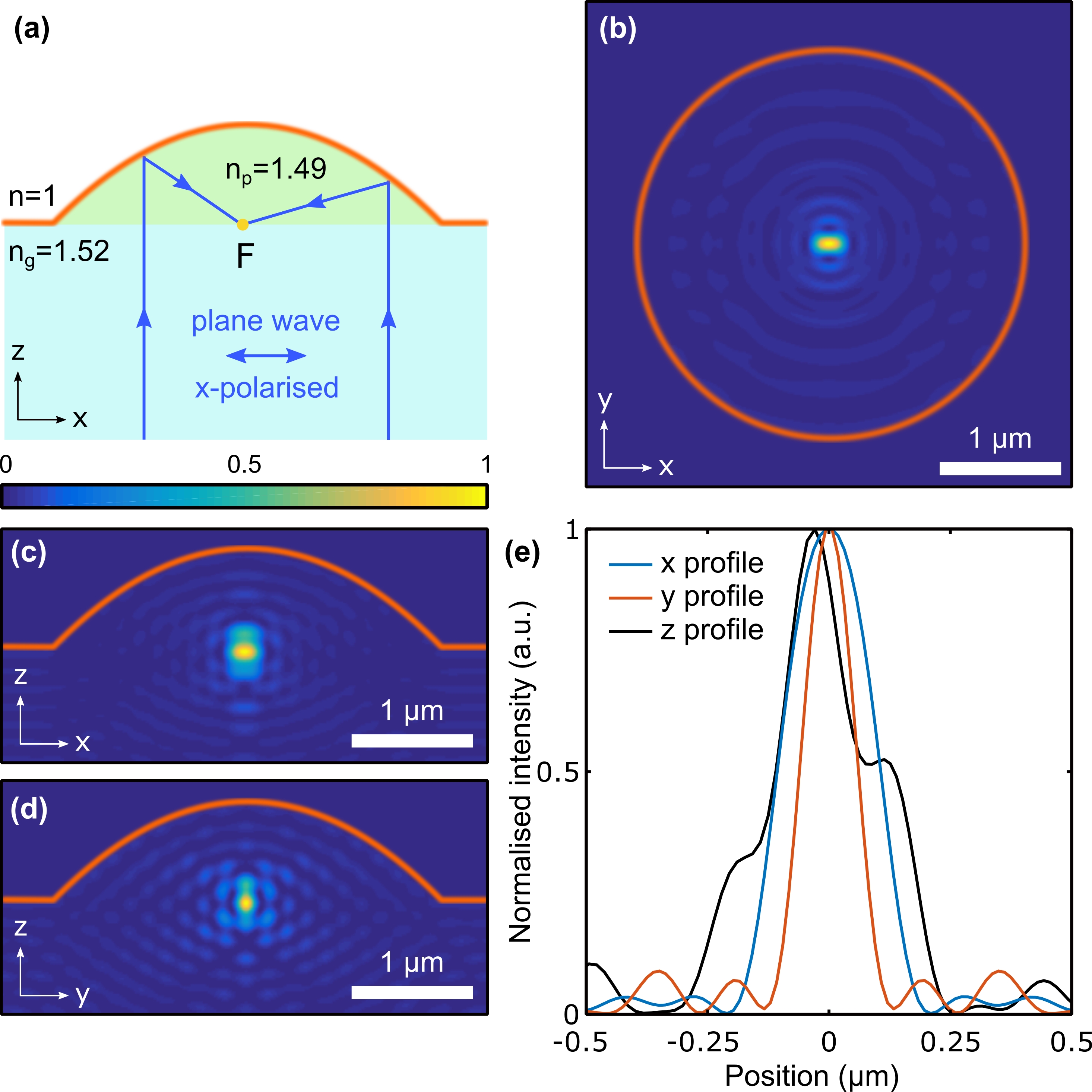}
\caption{\label{fig:intro} Micro parabolic mirror for focusing light at the nanoscale.
	(a)~In geometrical optics, a parabolic mirror concentrates the collimated illumination in its focal point $F$. 
	(b-d)~FDTD simulations of intensity distribution in a micro parabolic mirror with focal length of 0.8~$\mu$m illuminated with a collimated $x$-polarised plane wave at $\lambda=442$~nm. 
	The intensity distributions of total electric field intensity are shown in $x-y$ focal plane in (b),  as well as in orthogonal $x-z$ and $y-z$ planes through the focal point in (c) and (d), respectively. The outline of parabolic mirror is shown by orange line.
	(e)~Sections of the intensity profiles from panels (b-d) demonstrate the sub-wavelength focal spot of the micro parabolic mirror.}
\end{figure}

We consider a micro parabolic mirror with focal length  $f=0.8~\mu$m and aperture diameter $d=3.2~\mu$m  covering a $2\pi$ solid angle.  {In such a geometry, the radius $R$ of the dish is double of its height $H$, that is $R=2H=2f$.}
The dish is filled with a polymer ($n_{p}=1.49$) and illuminated from a glass substrate ($n_{g}=1.52$) with a linearly polarised plane wave at $\lambda_{ex}=442$~nm as sketched in  Fig.\ref{fig:intro}(a). 
Due to the small size of the mirror, diffraction effects are expected, which can be well captured in finite-difference time-domain (FDTD) numerical simulations. 
The map of total electric field intensity $|\bm{E}|^2$ through the focal spot is plotted in Fig.\ref{fig:intro}(b-d). 
The focal plane in Fig.\ref{fig:intro}(b) demonstrates that the maximum confinement is achieved in the $y$ direction, which is orthogonal to the excitation polarisation ($x$ here). 

The intensity cross-sections through the focal spot in $x$, $y$, and $z$ directions are plotted in Fig.\ref{fig:intro}(e), where the maximum of normalised total electric field intensity in  $z$ direction is  close to the position of geometrical focal point, just 30~nm shifted into the glass substrate. 
We ascribe this to the small refractive index mismatch of the parabolic mirror filling $n_{p}$ and the glass substrate $n_{g}$ (see SI Fig.S1). 
The obtained intensity distributions around the focal point have a full width at half maximum (FWHM) in lateral ($x$ and $y$) and axial ($z$) directions of $\Delta x = 222$~nm, $\Delta y = 120$~nm and $\Delta z = 247$~nm. 
The focal spot size scales with the refractive index of the parabolic mirror filling, and thus a much tighter focusing can be reached in comparison with an air-filled parabolic mirror (see SI Fig.S2).   
Hence, the sub-wavelength focus of $\sim \lambda_{ex}/4$ can be achieved for a focal length of only 0.8~$\mu$m, being unaffected by size-effects and  diffraction. 
 Mirrors with  focal length shorter than that   drastically decrease the intensity in the focal point (see SI Fig.S3), and loose the ability to collimate the generated emission \cite{Morozov2018}. 

Another important characteristic of high-NA parabolic mirrors is their ability to convert the incident light polarisation from transverse ($x$ here) to longitudinal ($z$ here), allowing for the generation of optical fields  with a strong electric field component along the optical axis \cite{Drechsler2001,Debus2003}. 
The numerical simulations confirm the polarisation conversion,  {indicating a $|\bm{E_x}|^2$ component is the dominant contribution to the intensity in focal region,} while $|\bm{E_y}|^2$ and $|\bm{E_z}|^2$ components have 24 and 4 times lower magnitude (see SI Fig.S4).  
Hence, the micro parabolic mirror is able to excite also quantum emitters with electric dipoles oriented perpendicularly to the sample plane, with about  {$20$\%}  efficiency. 

\begin{figure*}[tbh!]
    \includegraphics[width=1\linewidth]{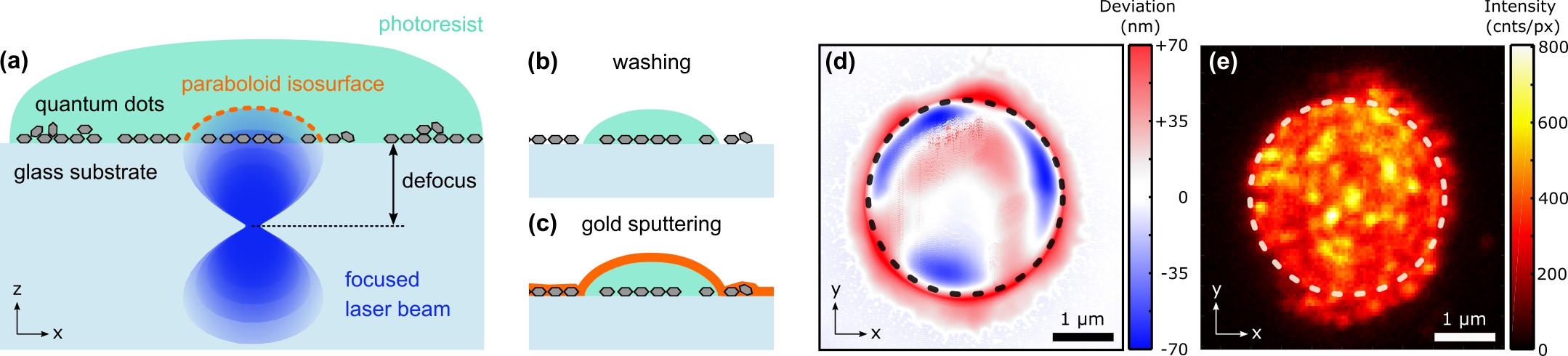}
	\caption{\label{fig:fabrication} Fabrication of micro parabolic mirrors by in-situ volumetric lithography.
	(a)~The laser beam was focused below the sample plane to polymerise a parabolic micro structure over a quantum dot layer (not to scale). (b) The unexposed photoresist was washed away resulting in a polymeric parabolic dish on glass substrate. (c)~The sample was coated with 80~nm gold layer  to form a micro parabolic mirror with quantum dots in its focal plane. 
	(d)~Deviation of a micro parabolic mirror surface from a perfect paraboloid shape was obtained by AFM scanning. The black dashed circle represents the mirror aperture. 
	(e)~Confocal scan of the parabolic mirror focal plane reveals a quasi homogeneous photoluminescence intensity distribution, while the photoluminescence of quantum dots outside aperture (dashed white circle) is completely quenched by the deposited gold layer.
	}
\end{figure*}

\subsection{Fabrication}
We fabricated  micro parabolic mirrors over a fluorescent layer composed of colloidal giant shell CdSe/CdS quantum dots emitting at $\lambda_{em}=650$~nm \cite{Christodoulou2014}.  {The giant shell configuration of quantum dots is highly photostable, and helped us to eliminate such undesirable artifacts as photobleaching.}
The quantum dots are used here to map the intensity distribution in the focal plane of the micro parabolic mirror, as their fluorescence is proportional to the excitation light intensity, provided that it is well below the saturation intensity. 
A layer of quantum dots was deposited  by  spin-coating over a glass coverslip prior to the fabrication  {of the mirror dish. 
In the fabrication, we followed the one-photon direct laser writing which we have described in \cite{Morozov2018}. 
Such one-photon photopolymerization has a power threshold \cite{Delrot2018}, thus a clear boundary in the polymerization is formed. 
Our lithography technique is based on the polymerization of structures with the volumetric intensity profile of a focused Gaussian beam (the blue hourglass-shaped  profile in Fig.\ref{fig:fabrication}).} 
First, we localised the quantum dots by means of scanning confocal microscopy retrieving the position of the sample plane. 
Next, we  {controllably defocused the confocal system using a piezo stage in order to expose the photoresist to a part of hourglass-shaped intensity iso-surface (Fig.\ref{fig:fabrication}(a)). 
This} outer part of intensity iso-surface follows a paraboloid shape allowing for the polymerisation of parabolic structures in a single laser exposure step.  
After  washing off the unexposed photoresist, a parabolic polymer structure reveals over the quantum dot layer (Fig.\ref{fig:fabrication}(b)).  
In the final step, we sputtered an 80~nm gold layer over the sample to form a metal parabolic mirror (Fig.\ref{fig:fabrication}(c)).  
The fabrication technique results in a smooth mirror surface of  high optical quality as demonstrated in Fig.\ref{fig:fabrication}(d): the deviation of a micro parabolic mirror shape from a perfect paraboloid surface was within $\pm70$~nm, extracted from a 3D fit of an atomic force microscope (AFM) scan. 
A confocal scan in Fig.\ref{fig:fabrication}(e) shows the focal plane of a fabricated parabolic mirror, where the fluorescence signal originates from the quantum dot layer. 
The fluorescence signal  within the parabolic mirror aperture is distributed quasi homogeneously without a dominant intensity spot ($I_{avg}=392$~cnts/px,  $I_{std}=96$~cnts/px), while the quantum dot layer fluorescence outside the mirror aperture is completely quenched by the deposited gold layer. 

\subsection{Excitation of quantum emitters}

\begin{figure}
    \includegraphics[width=1\linewidth]{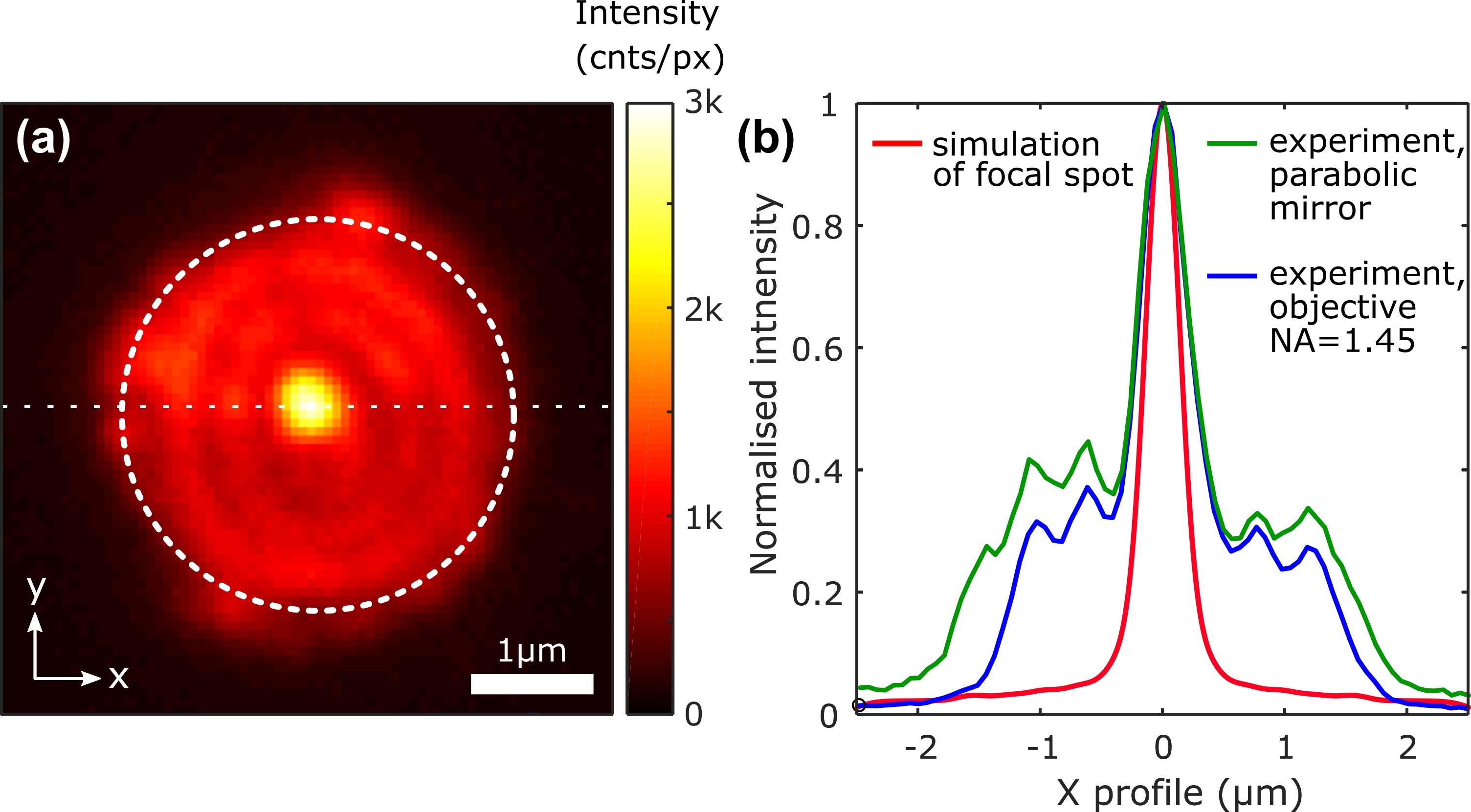}
	\caption{\label{fig:excitation} Excitation of quantum emitters with the micro parabolic mirror.
	(a)~A~collimated laser beam at $\lambda_{ex}=442$~nm causes fluorescence of all quantum dots in the focal plane, while the parabolic mirror reflects the laser beam exciting quantum dots only in its focal spot resulting in a bright intensity spot in the centre of mirror aperture (dashed white circle). (b)~The green intensity cross section of the micro parabolic focal spot was acquired along the dotted white  line in panel (a). The blue cross section was obtained by focusing with the high NA objective in the position of parabolic mirror focal spot. The red line shows the simulated focal spot intensity cross section.  
	}	
\end{figure}

The fabricated micro parabolic mirror with a quantum dot fluorescence layer in its focal plane was illuminated with a collimated laser beam at $\lambda_{ex}=442$~nm to demonstrate its focusing properties. The resulting fluorescence intensity profile of the parabolic mirror focal plane (Fig.\ref{fig:excitation}(a)) was imaged on a CCD camera using an objective (Nikon Plan Apochromat 100x, NA = 1.45). The obtained intensity distribution in Fig.\ref{fig:excitation}(a) is very different from the confocal intensity map presented in Fig.\ref{fig:fabrication}(e), and a dominant intensity peak is clearly visible in the position of the micro parabolic mirror focal spot. The residual background in Fig.\ref{fig:excitation}(b) comes from fluorescence signal of the quantum dots in the focal plane of the micro parabolic mirror excited by the collimated laser beam. In addition, the intensity profile is modulated by interference rings around the central bright focal spot, which we attribute to  the reflection of photoluminescence between parabolic  surface and the focal plane (see SI Fig.S7). 

In order to characterise the fluorescence intensity distribution in the focal plane of the parabolic mirror (Fig.\ref{fig:excitation}(a)), we extract an intensity cross section through the focal spot as shown by green line in Fig.\ref{fig:excitation}(b). The experimental intensity cross section is characterised by a FWHM of $\Delta_{exp}=554\pm38$~nm, which is comparable with the simulated focal spot size $\Delta_{sim}=347$~nm. This simulated focal spot shown by red line in Fig.\ref{fig:excitation}(b) was obtained by a convolution of the parabolic mirror excitation profile presented in Fig.\ref{fig:intro} with the point spread function $\Delta_{PSF}=224$~nm of the imaging system with NA=1.45 at $\lambda_{em}=650$~nm, estimated from the Abbe diffraction limit (see SI Fig.S5). The difference in FWHM between the experimental and simulated values of focal spot sizes is due to the presence of the quantum dot layer, which can scatter both the exciting beam and the emitted light, thus slightly widening the collected spot. 

To compare the performance of the parabolic mirror we also directly excited the quantum dots layer at the position of the micro parabolic mirror focal spot with the high NA objective. In  Fig.\ref{fig:excitation}(b) we present a cross section (blue line) of the photoluminescence intensity excited in this way (see also SI Fig.S6). This profile has the same FWHM of $\Delta_{exp}=555\pm38$~nm as in the case of focusing with the micro parabolic mirror. While one could conclude that the micro parabolic mirror focusing power is comparable to that of the 1.45 NA objective,  {we point out that the measured spot size comes from (i) excitation through the mirror with an expected average focal spot size of $\Delta_{par}$=171~nm (Fig.\ref{fig:intro}), and (ii) imaging through an objective with finite resolution. Although we expect a slightly smaller focal spot size for a 1.45 NA objective ($\Delta_{obj}$=152~nm), its measured focal spot would be very similar to the micro parabolic mirror (see SI). Therefore, we conclude that the results shown in Fig.\ref{fig:excitation}(b)  are compatible with the expected focusing power of the micro parabolic mirror, and $\Delta_{exp}=555$~nm is only an upper bound for the size of the focal spot.}

\subsection{Directing quantum emission}

\begin{figure}
    \includegraphics[width=1\linewidth]{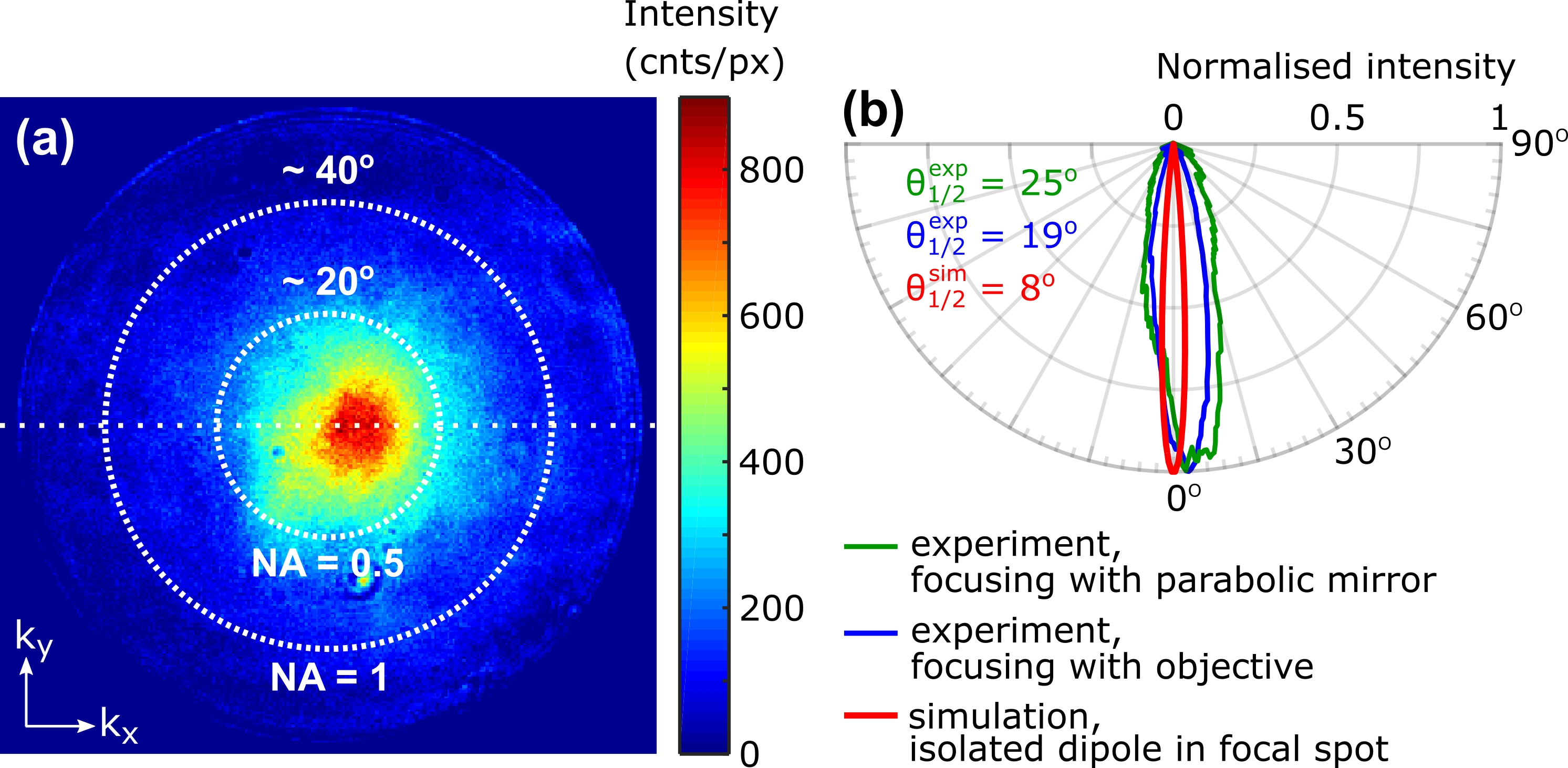}
 	\caption{\label{fig:directing} The fluorescence of quantum dots in the focal spot of the micro parabolic mirror is collimated.
	    (a)~A~Fourier space image of quantum dots in the focal spot of the micro parabolic mirror presented in Fig.\ref{fig:excitation}(a). The main lobe of the radiation pattern is completely within NA=0.5. (b)~Radiation pattern in polar coordinates demonstrate the low beam divergence. The radiation pattern obtained using the micro parabolic antenna excitation is shown by green curve acquired along the dotted white line in the panel (a). The blue line represents the radiation pattern obtained using confocal excitation of the quantum dots in the focal spot of the same micro parabolic mirror.  The red line is the simulation  of an isolated single  {$x$-oriented} dipole in the focal spot of the micro parabolic mirror.  
	    }	
\end{figure}

The photons emitted by the quantum dots in the focal spot are collimated by the micro parabolic mirror. We demonstrate this by imaging the back focal plane of microscope objective to obtain the Fourier space momentum distribution of quantum dot emission. Such a Fourier space image of the micro parabolic mirror focal plane intensity profile from Fig.\ref{fig:excitation}(a) is presented  in Fig.\ref{fig:directing}(a).  The emission of the quantum dots excited with the micro parabolic mirror is directed into a narrow beam along the optical axis, which is completely contained in an NA=0.5 solid angle. The fluorescence beam collimation is characterised in Fig.\ref{fig:directing}(b), which presents the intensity cross sections of the micro parabolic mirror radiation pattern.  {
This result summarizes the full operational principle of the parabolic mirror: first it efficiently couples a directional plane wave onto the emitter, similarly to what achieved with a conventional objective \cite{Morozov2018}, and then it directs the generated photoluminescence back to the far-field. 
}

The simulation of an isolated single horizontal dipole in the focal spot of the micro parabolic mirror confirms the unidirectional radiation pattern and is plotted as a red curve in Fig.\ref{fig:directing}(b), with a half power beam width of $\theta_{1/2}^{sim}=8^\circ$ at $\lambda_{em}=650$~nm.  The green curve in Fig.\ref{fig:directing}(b) is the radiation pattern cross section obtained from Fig.\ref{fig:directing}(a), and the micro parabolic mirror excitation with a collimated laser beam in the widefield configuration yields in the collimated fluorescence beam with $\theta_{1/2}^{exp}=25^\circ$ (after the background subtraction, see SI Fig.S8). The difference in the half power beam width originates from the layer of quantum dots distributed quasi-homogeneously  in the sample plane. The emission of the quantum dots out of the focal spot of the micro parabolic mirror is directional as well, however it is not aligned with the mirror optical axis. The parabolic mirror reflects their emission at different angles, and thus broadens the radiation pattern in Fig.\ref{fig:directing}. The background would be absent in case of a single quantum dot in the focal point of the micro parabolic mirror as we have shown in ref.~\cite{Morozov2018}. In the current experimental conditions the background contribution can be reduced by confocal excitation with the high NA objective in the position of focal spot of micro parabolic antenna resulting in the collimated emission with $\theta_{1/2}^{exp}=19^\circ$ (blue curve in Fig.\ref{fig:directing}(b)).

\section{Conclusion}

In conclusion, we scaled down to microscale the concept of a reflective objective based on the micro parabolic mirror. We fabricated such a compact parabolic mirror with a focal length of only $0.8$~$\mu$m, capable of focusing the excitation light to sub-wavelength spot and to extract the fluorescence from nanoscale emitters. The fabrication process is fast, single-shot, and leads to a low surface roughness.
The volumetric lithography provides fast fabrication of microstructures of high optical quality in sub second single laser shots over large areas, which can be also aligned with individual emitters. 
This is an ideal approach to objective-free microscopy, especially for cryogenic and vacuum conditions, and could become a powerful tool in nanoscale quantum optics.

The presented design of the micro parabolic mirror covers  2$\pi$ solid angle around an emitter in the focal point, which limits the focal spot size as well as the photon extraction efficiency. The mirror directs photons out of the sample plane, while more sophisticated designs could be exploited for applications where photons need to be guided in plane.
The in-situ volumetric lithography allows for fabrication of structures beyond the parabolic shape at the microscale. More complex spatial intensity distribution could be achieved by using wavefront-shaping techniques, such as spatial light modulators (SLMs) or digital micromirror devices (DMDs) \cite{Jenness2010,Tian2020}.

\section*{Supporting Information}
See the Supporting Information for the effect of refractive index mismatch between parabolic mirror filling and glass substrate; scaling of the focal spot size with the refractive index of parabolic dish filling; evolution of the focal spot dimensions with further decreasing of parabolic mirror focal length; polarization of the parabolic mirror focal spot; simulation of experimental intensity distribution in the focal point; interference rings in the focal plane of parabolic mirror; measurements and simulations of radiation pattern.

\begin{acknowledgments}
S.M. and R.S. acknowledge funding by EPSRC (EP/P033369 and EP/M013812/1). A.H.K. and I.M. acknowledge funding from the European Research Council (ERC) under the European Unions Horizon 2020 research and innovation program (714876 PHOCONA).
\end{acknowledgments}

\section*{Data availability}
The data that support the findings of this study are openly available in Figshare at http://doi.org/10.6084/m9.figshare.12357356.

\bibliography{aipsamp}

\end{document}